\documentstyle[epsf,twoside]{article}

\catcode`\@=11
\long\def\@makefntext#1{
\protect\noindent \hbox to 3.2pt {\hskip-.9pt  
$^{{\eightrm\@thefnmark}}$\hfil}#1\hfill}		

\def\@makefnmark{\hbox to 0pt{$^{\@thefnmark}$\hss}}	
	
\def\ps@myheadings{\let\@mkboth\@gobbletwo
\def\@oddhead{\hbox{}
\rightmark\hfil\eightrm\thepage}   
\def\@oddfoot{}\def\@evenhead{\eightrm\thepage\hfil
\leftmark\hbox{}}\def\@evenfoot{}
\def\sectionmark##1{}\def\subsectionmark##1{}}

\oddsidemargin=\evensidemargin
\newcounter{sectionc}\newcounter{subsectionc}\newcounter{subsubsectionc}
\renewcommand{\section}[1] {\vspace{12pt}\addtocounter{sectionc}{1} 
\setcounter{subsectionc}{0}\setcounter{subsubsectionc}{0}\noindent 
	{\tenbf\thesectionc. #1}\par\vspace{5pt}}
\renewcommand{\subsection}[1] {\vspace{12pt}\addtocounter{subsectionc}{1} 
	\setcounter{subsubsectionc}{0}\noindent 
	{\bf\thesectionc.\thesubsectionc. {\kern1pt \bfit #1}}\par\vspace{5pt}}
\renewcommand{\subsubsection}[1] {\vspace{12pt}\addtocounter{subsubsectionc}{1}
	\noindent{\tenrm\thesectionc.\thesubsectionc.\thesubsubsectionc.
	{\kern1pt \tenit #1}}\par\vspace{5pt}}

\newcounter{appendixc}
\newcounter{subappendixc}[appendixc]
\newcounter{subsubappendixc}[subappendixc]
\renewcommand{\thesubappendixc}{\Alph{appendixc}.\arabic{subappendixc}}
\renewcommand{\thesubsubappendixc}
	{\Alph{appendixc}.\arabic{subappendixc}.\arabic{subsubappendixc}}

\renewcommand{\appendix}[1] {\vspace{12pt}
        \refstepcounter{appendixc}
        \setcounter{figure}{0}
        \setcounter{table}{0}
        \setcounter{lemma}{0}
        \setcounter{theorem}{0}
        \setcounter{corollary}{0}
        \setcounter{definition}{0}
        \setcounter{equation}{0}
        \renewcommand{\thefigure}{\Alph{appendixc}.\arabic{figure}}
        \renewcommand{\thetable}{\Alph{appendixc}.\arabic{table}}
        \renewcommand{\theappendixc}{\Alph{appendixc}}
        \renewcommand{\thelemma}{\Alph{appendixc}.\arabic{lemma}}
        \renewcommand{\thetheorem}{\Alph{appendixc}.\arabic{theorem}}
        \renewcommand{\thedefinition}{\Alph{appendixc}.\arabic{definition}}
        \renewcommand{\thecorollary}{\Alph{appendixc}.\arabic{corollary}}
        \renewcommand{\theequation}{\Alph{appendixc}.\arabic{equation}}
        \noindent{\tenbf Appendix \theappendixc #1}\par\vspace{5pt}}
\newcommand{\subappendix}[1] {\vspace{12pt}
        \refstepcounter{subappendixc}
        \noindent{\bf Appendix \thesubappendixc. {\kern1pt \bfit #1}}
	\par\vspace{5pt}}
\newcommand{\subsubappendix}[1] {\vspace{12pt}
        \refstepcounter{subsubappendixc}
        \noindent{\rm Appendix \thesubsubappendixc. {\kern1pt \tenit #1}}
	\par\vspace{5pt}}

\newcommand{\textlineskip}{\baselineskip=13pt}
\newcommand{\smalllineskip}{\baselineskip=10pt}

\def\eightcirc{
\begin{picture}(0,0)
\put(4.4,1.8){\circle{6.5}}
\end{picture}}
\def\eightcopyright{\eightcirc\kern2.7pt\hbox{\eightrm c}} 

\newcommand{\copyrightheading}[1]
	{\vspace*{-2.5cm}\smalllineskip{\flushleft
	{\footnotesize International Journal of Modern Physics A, #1}\\
	{\footnotesize $\eightcopyright$\, World Scientific Publishing
	 Company}\\
	 }}

\def\abstracts#1#2#3{{
	\centering{\begin{minipage}{4.5in}\baselineskip=10pt\footnotesize
	\parindent=0pt #1\par 
	\parindent=15pt #2\par
	\parindent=15pt #3
	\end{minipage}}\par}} 

\newcounter{itemlistc}
\newcounter{romanlistc}
\newcounter{alphlistc}
\newcounter{arabiclistc}

\newcommand{\fcaption}[1]{
        \refstepcounter{figure}
        \setbox\@tempboxa = \hbox{\footnotesize Fig.~\thefigure. #1}
        \ifdim \wd\@tempboxa > 5in
           {\begin{center}
        \parbox{5in}{\footnotesize\smalllineskip Fig.~\thefigure. #1}
            \end{center}}
        \else
             {\begin{center}
             {\footnotesize Fig.~\thefigure. #1}
              \end{center}}
        \fi}

\newcommand{\tcaption}[1]{
        \refstepcounter{table}
        \setbox\@tempboxa = \hbox{\footnotesize Table~\thetable. #1}
        \ifdim \wd\@tempboxa > 5in
           {\begin{center}
        \parbox{5in}{\footnotesize\smalllineskip Table~\thetable. #1}
            \end{center}}
        \else
             {\begin{center}
             {\footnotesize Table~\thetable. #1}
              \end{center}}
        \fi}

\def\@citex[#1]#2{\if@filesw\immediate\write\@auxout
	{\string\citation{#2}}\fi
\def\@citea{}\@cite{\@for\@citeb:=#2\do
	{\@citea\def\@citea{,}\@ifundefined
	{b@\@citeb}{{\bf ?}\@warning
	{Citation `\@citeb' on page \thepage \space undefined}}
	{\csname b@\@citeb\endcsname}}}{#1}}

\newif\if@cghi
\def\cite{\@cghitrue\@ifnextchar [{\@tempswatrue
	\@citex}{\@tempswafalse\@citex[]}}
\def\citelow{\@cghifalse\@ifnextchar [{\@tempswatrue
	\@citex}{\@tempswafalse\@citex[]}}
\def\@cite#1#2{{$\null^{#1}$\if@tempswa\typeout
	{IJCGA warning: optional citation argument 
	ignored: `#2'} \fi}}

\def\pmb#1{\setbox0=\hbox{#1}
	\kern-.025em\copy0\kern-\wd0
	\kern.05em\copy0\kern-\wd0
	\kern-.025em\raise.0433em\box0}

\def\fnt#1#2{\footnotetext{\kern-.3em
	{$^{\mbox{\scriptsize #1}}$}{#2}}}

\def\fpage#1{\begingroup
\voffset=.3in
\thispagestyle{empty}\begin{table}[b]\centerline{\footnotesize #1}
	\end{table}\endgroup}

\def\runninghead#1#2{\pagestyle{myheadings}
\markboth{{\protect\footnotesize\it{\quad #1}}\hfill}
{\hfill{\protect\footnotesize\it{#2\quad}}}}
\font\tenrm=cmr10
\font\tenit=cmti10 
\font\tenbf=cmbx10
\font\bfit=cmbxti10 at 10pt
\font\ninerm=cmr9

\font\eightrm=cmr8

\def\qed{\hbox{${\vcenter{\vbox{			
   \hrule height 0.4pt\hbox{\vrule width 0.4pt height 6pt
   \kern5pt\vrule width 0.4pt}\hrule height 0.4pt}}}$}}

\newcommand{\nub}{\overline{\nu}}
\newcommand{\cbar}{\overline{c}}

\begin{document}

\runninghead{New Measurements of Nucleon Structure Functions
Manuscripts $\ldots$} {New Measurements of Nucleon Structure Functions
Manuscripts $\ldots$}

\normalsize\textlineskip
\thispagestyle{empty}
\setcounter{page}{1}

\copyrightheading{}			

\vspace*{0.88truein}

\fpage{1}
\centerline{\bf NEW MESUREMENTS OF NUCLEON STRUCTURE}
\vspace*{0.035truein}
\centerline{\bf FUNCTIONS FROM CCFR/NuTeV }
\vspace*{0.37truein}
\noindent
{\footnotesize A.~Bodek,$^{7}$ U.~K.~Yang,$^{7}$ T.~Adams, $^{4}$ A.~Alton,$^{4}$
C.~G.~Arroyo,$^{2}$ S.~Avvakumov,$^{7}$ L.~de~Barbaro,$^{5}$
P.~de~Barbaro, $^{7}$ A.~O.~Bazarko,$^{2}$ R.~H.~Bernstein,$^{3}$
 T.~Bolton, $^{4}$ J.~Brau,$^{6}$ D.~Buchholz,$^{5}$
H.~Budd,$^{7}$ L.~Bugel,$^{3}$ J.~Conrad,$^{2}$ R.~B.~Drucker,$^{6}$
B.~T.~Fleming,$^{2}$
J.~A.~Formaggio,$^{2}$ R.~Frey,$^{6}$ J.~Goldman,$^{4}$
M.~Goncharov,$^{4}$ D.~A.~Harris,$ ^{7} $ R.~A.~Johnson,$^{1}$
J.~H.~Kim,$^{2}$ B.~J.~King,$^{2}$ T.~Kinnel,$ ^{8}$
S.~Koutsoliotas,$^{2}$ M.~J.~Lamm,$^{3}$ W.~Marsh,$^{3}$
D.~Mason,$^{6}$ K.~S.~McFarland, $^{7}$ C.~McNulty,$^{2}$
S.~R.~Mishra,$^{2}$ D.~Naples,$^{4}$ N.~Suwonjandee,$^{1}$ P.~Nienaber,$^{3}$
A.~Romosan,$^{2}$ W.~K.~Sakumoto,$^{7}$ H. Schellman,$^{5}$
F.~J.~Sciulli,$^{2}$ W.~G.~Seligman,$^{2}$ M.~H.~Shaevitz,$^{2}$
W.~H.~Smith,$^{8}$ P.~Spentzouris,$^{2}$ E.~G.~Stern,$^{2}$
M.~Vakili,$^{1}$ A.~Vaitaitis,$^{2}$  
J.~Yu,$^{3}$ G.~P.~Zeller,$^{5}$ and E.~D.~Zimmerman$^{2}$}\\
\vspace*{0.1truein}
\centerline{\footnotesize (Presented at DPF2000 by Arie Bodek for the CCFR/NuTeV  
Collaboration - UR-1615)}
\vspace*{0.015truein}
\centerline{\footnotesize\it $^{1}$ Univ. of Cincinnati, Cincinnati, OH 45221;
$^{2}$ Columbia University, New York, NY 10027} 
\vspace*{0.015truein}
\centerline{\footnotesize\it $^{3}$ Fermilab, Batavia, IL 60510
$^{4}$ Kansas State University, Manhattan, KS 66506}
\vspace*{0.015truein}
\centerline{\footnotesize\it$^{5}$ Northwestern University, Evanston, IL 60208;
$^{6}$ Univ. of Oregon, Eugene, OR 97403}
\vspace*{0.015truein}
\centerline{\footnotesize\it $^{7}$ Univ. of Rochester, Rochester, NY 14627;
$^{8}$ Univ. of Wisconsin, Madison, WI 53706}


\vspace*{0.21truein}
\abstracts{
We report on the extraction of  the structure functions
$F_2$ and  $\Delta xF_3 = xF_3^{\nu}-xF_3^{\nub}$
from CCFR  $\nu_\mu$-Fe and $\nub_\mu$-Fe differential cross sections.
The extraction is performed in a physics model independent (PMI) way.
This first measurement for  $\Delta xF_3$, which is useful in testing
models of heavy charm production, is higher
than current theoretical predictions. Within 5$\%$
the $F_2$ (PMI) values  measured in $\nu_\mu$ and $\mu$ scattering
are in agreement with the predictions of
Next-to-Leading-Order PDFs (using massive charm production 
schemes), thus resolving the long-standing discrepancy
between the two measurements.
}{}{}

\textlineskip			
\vspace*{12pt}			

\noindent

%
Deep inelastic lepton-nucleon scattering experiments have been used
to determine the quark distributions in the nucleon.
However, the quark distributions determined from $\mu$
and $\nu$
experiments were found to be different at small values of $x$,
because of a disagreement in the extracted structure functions.
Here,
we report on a measurement of differential cross sections
and structure functions from CCFR $\nu_\mu$-Fe and $\nub_\mu$-Fe data.
We find that the neutrino-muon difference is resolved by
extracting the $\nu_\mu$ structure functions in a physics
model independent way.

The sum of $\nu_\mu$ and $\nub_\mu$
 differential cross sections 
for charged current interactions on an isoscalar target is related to the
structure functions as follows:
\begin{tabbing}
$F(\epsilon)$ \= $\equiv \left[\frac{d^2\sigma^{\nu }}{dxdy}+
\frac{d^2\sigma^{\overline \nu}}{dxdy} \right] 
 \frac {(1-\epsilon)\pi}{y^2G_F^2ME_\nu}$
 $ = 2xF_1 [ 1+\epsilon R ] + \frac {y(1-y/2)}{1+(1-y)^2} \Delta xF_3 $.
\end{tabbing}
Here $G_{F}$ is the Fermi weak coupling constant, $M$ is the nucleon
mass, $E_{\nu}$ is the incident energy, the scaling
variable $y=E_h/E_\nu$ is the fractional energy transferred to
the hadronic vertex, $E_h$ is the final state hadronic
energy, and $\epsilon\simeq2(1-y)/(1+(1-y)^2)$ is the polarization of 
the virtual $W$ boson.
The structure
function $2xF_1$ is expressed in terms of $F_2$
by $2xF_1(x,Q^2)=F_2(x,Q^2)\times
\frac{1+4M^2x^2/Q^2}{1+R(x,Q^2)}$, where $Q^2$ is the
square of the four-momentum transfer to the nucleon,
  $x=Q^2/2ME_h$ (the Bjorken scaling variable) is
the fractional momentum carried by the struck quark,
and $R=\frac{\sigma
_{L}}{\sigma _{T}} $ is the ratio of the cross-sections of
longitudinally- to transversely-polarized $W$ bosons.
The $\Delta xF_3$ term, which in leading order
 $\simeq 4x(s-c)$, is not present in the $\mu$-scattering case.
In addition, in a  $\nu_\mu$ charged current interaction with
$s$ (or $\cbar$) quarks, there is a threshold suppression originating
from the production of heavy $c$ quarks in the final state.
For $\mu$-scattering, there is no suppression for scattering
from $s$ quarks, but more suppression when scattering
from $c$ quarks since there are two heavy
 quarks ($c$ and $\cbar$) in the final state.  

In previous analyses
of $\nu_\mu$ data,
 structure functions were extracted
by applying a slow rescaling correction to correct for the charm
mass suppression in the final state. In addition, 
the $\Delta xF_3$ term (used as input in the extraction)
was calculated from a leading order charm production model. These
resulted in a physics model dependent (PMD) structure functions. 
 In the
new analysis reported here, slow rescaling corrections are not applied,
 and $\Delta xF_3$ and $F_2$
are extracted from two-parameter fits to the data.
We compare the values of $\Delta xF_3$
to various charm production models.
The extracted physics model independent (PMI) values for  $F_2^{\nu}$
are then compared  with $F_2^{\mu}$
within the framework of NLO models for massive charm production.

\begin{figure}[htbp]
\centerline{\mbox{\epsfxsize=3.37in \epsfysize=2.25in \epsffile{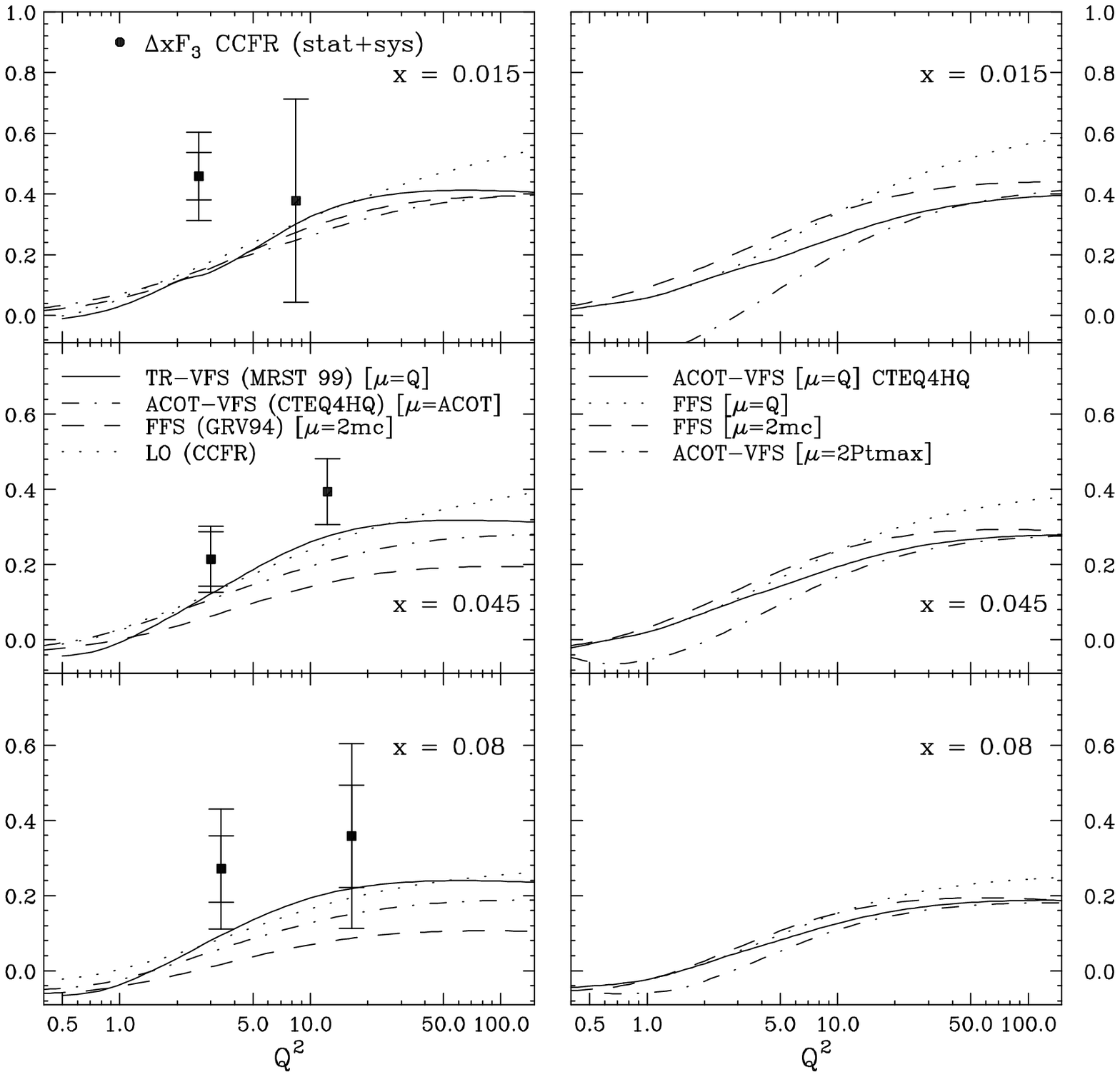}}}
\fcaption{
 $\Delta xF_3$ data as a function of $x$
compared with various schemes for massive charm production:
 RT-VFS(MRST),
 ACOT-VFS(CTEQ4HQ), FFS(GRV94), and LO(CCFR), a
leading order model with a slow rescaling correction (left);
Also shown is the sensitivity of the theoretical calculations
to the choice of scale (right).}
\label{fig:dxf3}
\end{figure}

Because of the limited statistics, we use 
large bins in $Q^2$ in the extraction of $\Delta xF_3$ with bin centering
corrections. 
Figure~\ref{fig:dxf3}   shows the extracted
values of $\Delta xF_3$ as a function of $x$, including both statistical
and systematic errors, compared to
various theoretical methods for modeling
heavy charm productions within a QCD framework. 
Figure~\ref{fig:dxf3} (right) also shows the sensitivity to
the choice of scale.
With reasonable choices of scale, all the theoretical models
yield similar results. However,
at low $Q^2$  our $\Delta xF_3$ data are higher than all
the theortical models.
The difference between data and theory
may be due to an underestimate
of the strange sea 
 at low $Q^2$, or from missing
NNLO terms.

\begin{figure}[htbp]
\centerline{\mbox{\epsfxsize=4.0in \epsfysize=3.45in \epsffile{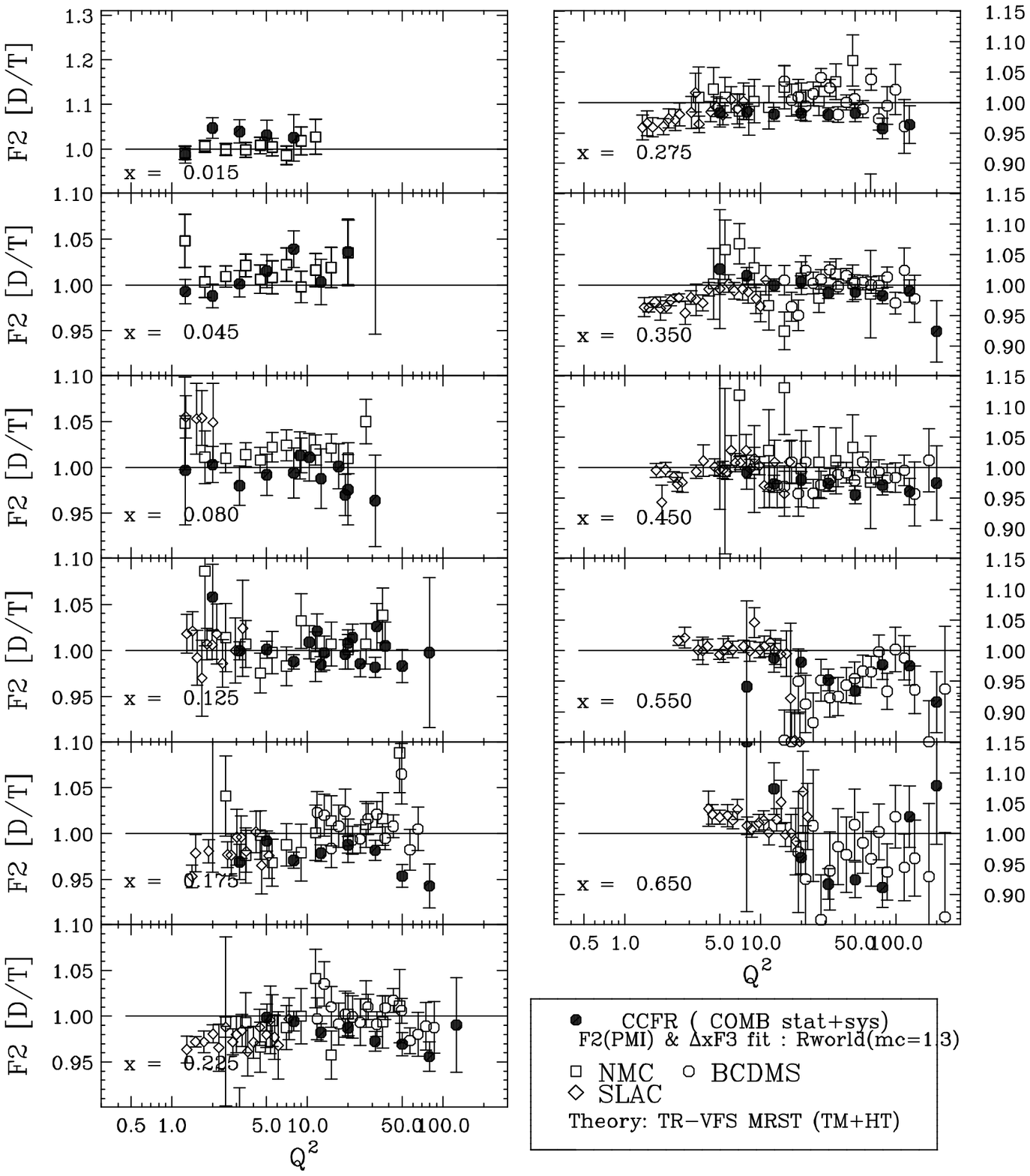}}}
\fcaption{
 The ratio (data/theory) of the $F_2^{\nu}$ (PMI)
data divided by the predictions of TR-VFS(MRST) (with
target mass and higher twist corrections).
Both statistical and systematic errors are included. Also shown
are the ratios of
the $F_2^{\mu}$ (NMC,BCDMS) and $F_2^e$ (SLAC) to the TR-VFS(MRST)
predictions. 
}
\label{fig:f2}
\end{figure}

Values of $F_2$ (PMI) for $x<0.1$ are extracted from
two-parameter fits to the $y$ distributions.
In the $x>0.1$ region,
the contribution from  $\Delta xF_3$ is small and the extracted
values of $F_2$ are insensitive to $\Delta xF_3$. Therefore,
we extract values of $F_2$ with an input value of $R$
and with $\Delta xF_3$ constrained to the TR-VFS(MRST) predictions. 
As in the case of the two-parameter fits for $x<0.1$, no 
corrections for slow rescaling are applied. 
Fig.~\ref{fig:f2} 
 shows our $F_2$ (PMI)
measurements divided by the predictions
from the TR-VFS(MRST) theory. Also shown are $F_2^{\mu}$ and
$F_2^e$ 
  divided
by the theory predictions. In the calculation of the QCD
 TR-VFS(MRST) predictions, we have also included corrections for nuclear
effects,
 target mass and
higher twist corrections
 at low values of $Q^2$. As seen in
Fig.~\ref{fig:f2}, within 5$\%$
both the neutrino and muon
structure functions are in agreement with the TR-VFS(MRST) predictions,
and therefore in agreement with each other. thus
resolving the long-standing discrepancy between the two
sets of data. A comparison
using the ACOT-VFS(CTEQ4HQ) predictions yields similar results.
Note that in the previous analysis
 of the CCFR data, the extracted values of $F_2$ (PMD) at the lowest
 $x=0.015$ and $Q^2$ bin
were up to $20\%$ higher than both the NMC data and
the predictions of the light-flavor
MRSR2 PDFs. More details on this work can be found
in UR- 1586 (hep-ex/0009041, submitted to Phys. Rev. Lett. 9/00),
and in U. K. Yang, Ph.D. Thesis, Univ. of Rochester
(UR-1583, in preparation).
\vspace*{12pt}
\end{document}